\documentclass{article}

\usepackage{arxiv}

\usepackage[utf8]{inputenc} 
\usepackage[T1]{fontenc}    
\usepackage{hyperref}       
\usepackage{url}            
\usepackage{booktabs}       
\usepackage{amsfonts}       
\usepackage{nicefrac}       
\usepackage{microtype}      
\usepackage{lipsum}
\usepackage{graphicx,verbatim,float,subfigure,graphicx}
\graphicspath{ {./images/} }

\title{Flexural Fatigue Life of Woven Carbon/Vinyl Ester Composites under Sea Water Saturation}

\author{
 Pavana Prabhakar \\
  Dept. of Civil \& Env. Engineering \\
  University of Wisconsin-Madison \\
  Madison, WI 53706 \\
  \texttt{pavana.prabhakar@wisc.edu} \\
\And
 Ricardo Garcia \\
  Dept. of Civil \& Env. Engineering \\
  University of Wisconsin-Madison \\
  Madison, WI 53706 \\
  \And
 Muhammad Imam \\
  Dept. of Civil \& Env. Engineering \\
  University of Wisconsin-Madison \\
  Madison, WI 53706 
  \AND
   Vinay Damodaran \\
  Dept. of Civil \& Env. Engineering \\
  University of Wisconsin-Madison \\
  Madison, WI 53706 \\
}

\date{ }

\begin{document}
\maketitle
\begin{abstract}
The adverse effects of sea water environment on the fatigue life of woven carbon fiber/vinyl ester composites are established at room temperature in view of long-term survivability of offshore structures. It is observed that the influence of sea water saturation on the fatigue life is more pronounced when the maximum cyclic displacement approaches maximum quasi-static deflection, that is, the reduction in the number of cycles to failure are comparable between dry and sea water saturated samples at lower strain ranges (~37\% at 0.46\% strain), but are drastically different at higher strain ranges (~90\% at 0.62\% strain). Key damage modes that manifest during the fatigue loading is also identified, and a non-linear model is established for predicting low cycle fatigue life of these composites in dry and sea water saturated conditions.
\end{abstract}


\section{Introduction}

Applications of fiber reinforced polymer composites (FRPCs) in various structural applications are rapidly increasing due to a sustained industrial demand for stronger and lighter components. FRPCs exhibit exceptional advantages over conventional materials such as high mechanical and thermal properties, temperature resistance, flexibility, and remarkable chemical and environmental resistance, in addition to high strength-to-weight ratio \cite{bortz2011,lars2010,tenney2009}. Structural designs which incorporate FRPCs for enhancing their mechanical load carrying capacity are subjected to a variety of loads, many of which are complex in nature and affect the mechanical properties of the material. This can lead to damage accumulation that may cause catastrophic failure of the structure. It is imperative for these load scenarios to be investigated thoroughly in order to understand their impact on composite structures, which may provide meaningful and useful data regarding the life-span of a structural component and enable the prediction of structural failure of a certain part or product. Environmental conditions play a critical role in determining the life-span and failure modes of a structure, and their influence on the load carrying capacity requires in-depth evaluation \cite{Davies2016,Noguchi1994,Staunton1982}. Often many structures are subjected to fluctuating and vibrating loads, typically categorized as fatigue loading, which are known to cause premature structural failure. This paper studies the influence of one particular load case, flexural cyclic loading (flexural fatigue), on woven carbon/vinyl ester composites under two different environmental conditions of dry and sea water saturation at room temperature.

Fatigue phenomenon is characterized by the failure caused by repeated loading, which initiate and propagate cracks as loading cycles increase. These types of loads can be steady, variable, uniaxial, or multiaxial. The cyclic (fatigue) load levels needed to cause failure is often less than the maximum quasi-static load, making this an important parameter to consider during design. Synthesis, analysis and testing are necessary procedures to develop a product with durability\cite{Stephens} for fatigue design as fatigue failures in structures implicate huge costs. Fatigue testing methods and design criteria for FRPCs can be challenging due to the complex damage mechanisms, which can potentially cause large scatter in fatigue life data and correspondingly increase the challenges associated with fatigue-predictive modeling. Temperature and environmental interactions with fatigue loading further generate intricate manifestations of failure modes within the material. Particularly, understanding the fatigue behavior of high-strength carbon fiber reinforced polymer (CFRP) composites has been of high importance for previous researchers \cite{REIFSNIDER1982187,Jones1984,BRUNBAUER201528,Curtis1989,BRUNBAUER201585}. In most of these research reported, fatigue studies have focused commonly on tension-tension (T-T), tension-compression (T-C) and compression-compression (C-C) loading at different stress ratios. Seldom research has been reported on the flexural fatigue performance of CFRP composites, and even slim on woven CFRP composites. Few studies\cite{NEWAZ1985,CAPRINO1998} exist on the flexural fatigue behavior of glass fiber reinforced polymer composites. A larger body of research exists on the quasi-static flexural behavior of CFRPs in different environmental conditions\cite{Bullions2003,Ishiaku2005,Nakamura2006,Hasnine2010,Mohanty2016,Garcia2019}. {\em To the best of the authors' knowledge, this is the first study on flexural fatigue behavior of woven CFRP composites under sea water saturation conditions.}

While fatigue is a topic widely investigated due to its frequent occurrence and relationship with structural failure, as mentioned above, seldom research has been reported for flexural (three-point bend) fatigue tests on woven carbon/vinyl ester composites in dry or moisture conditions at different temperatures. Other parameters such as test frequency, type of control (load or deflection), stress/strain ranges, and fixtures used contribute to the widespread types of fatigue tests in literature \cite{Curtis1989,REIFSNIDER1982187,DeBaere:2009}. In-service environment, most commonly moisture, temperature, UV radiation, chemical like sea water, etc. can have a significant influence on the mechanical performance and durability of FRPCs \cite{SELZER1997,Hartwig1984,Jiang2014,Xu2012}. These environments typically degrade the matrix material and the fiber/matrix interface, which are detrimental to their fatigue response. There has been limited work reported on the influence of temperature \cite{Kawai2001,AHLBORN1988} and moisture\cite{SELZER1997} on the fatigue response of CFRP composites.

As mentioned above, a majority of fatigue tests conducted on FRPCs have been performed in uniaxial tension/tension or tension/compression fatigue rather than in flexural fatigue \cite{GAMSTEDT:1999,Hansen:1999,Coats:1995,Caprino:2000,ROTEM:1991}. Nonetheless, scattered research is available on flexural fatigue combined with humidity-controlled settings and moisture uptake. Couillard and Schwartz\cite{Couillard:1997} studied the flexural fatigue behavior of composite materials featuring in-house cantilever beam bending fixtures, concluding that unidirectional carbon fiber/epoxy composite strands display fatigue damage as a loss of bending moment at 10$^6$  cycles. Chambers et al.\cite{Chambers:2006} investigated the effect of voids on the flexural fatigue performance of unidirectional carbon fiber composites, and showed compelling evidence that pre-existing voids drastically decrease the fatigue life of FRPCs.

The focus of the current work is on establishing the influence of sea water saturation on the flexural cyclic (fatigue) performance of woven carbon/vinyl ester composite samples at room temperature, in view of potential offshore marine applications. Vinyl ester resin is considered in the current study due to their superior UV resistance and lower water absorption as compared to polyester resins \cite{SOBRINHO:2009,Signor2003}, which makes it attractive for offshore marine applications. In addition, vinyl ester resins are widely used in sewer pipes, solvent storage tanks, automobile structural parts, swimming pools, building and construction, coating, and marine composites \cite{brown1997,MOURITZ1999,zhang2000,brody2005} due to their excellent chemical and corrosion resistance, favorable mechanical properties in addition to being low cost thermosets. Hence, the current study has a significant impact on a range of industries where vinyl esters are key resins in composites. The research presented in this paper has a widespread relevance to several industries.

\section{Materials and Methods}\label{Experimental Procedures}

\subsection{Manufacturing}
Carbon fiber laminates were fabricated using vacuum assisted resin transfer molding (VARTM) process. Material system, fabrication process and sample dimensions used in this study are discussed next.

\subsubsection{Material System}
3K tow (i.e. 3000 filaments per tow) plain weave carbon fiber fabrics purchased from FIBREGLAST {\textregistered} (\url{fibreglast.com}) were used to manufacture the laminates investigated in this paper. These fabrics create lightweight and tensile stiffened structural products, and are also compatible with a variety of thermosets and thermoplastics. This type of fabric is commonly used in aerospace, marine and automobile applications. Hetron 922 vinyl ester resin, formulated for 1.25\% MEKP, was the resin system used to impregnate the dry carbon fabric, which was also purchased from FIBREGLAST \textregistered. Hetron 922 is a low viscous thermoset, which is advantageous for easy infiltration during the VARTM process. Mechanical properties of the carbon fabric and vinyl ester resin are given in Table \ref{product}. 

\begin{table}[H]
\caption{Constituent material properties \cite{fibreglastFiber,fibreglastResin}}
  \centering
  \begin{tabular}{ccc}
    \hline Property & Carbon Fiber & Vinyl-Ester \\ \hline 			Tensile Strength & 4.2 - 4.4 GPa & 82 MPa \\  	Tensile Modulus & 227.5 - 240.6 GPa & 3.7 GPa \\  	Elongation & 1.4 - 1.95 \% & 4.6 - 7.9 \% \\  	Flexural Strength & - & 131 MPa \\  	Flexural Modulus & - & 3.4 GPa \\  	Nom. Thickness & 0.3048 mm & - \\ 	Barcol Hardness & - & 35 \\ \hline
  \end{tabular}
  \label{product}
\end{table}

\subsubsection{Laminate Fabrication}

VARTM process was used to fabricate composite panels with 305 $mm$ in length x 305 $mm$ in width. Carbon fabric along with additional textiles, such as flow media, breather and nylon sheets were also cut to fit aluminum molds of the same size as the expected composite panels. Material layers required for fabricating a single composite panel were as follows: 2 aluminum molds, 2 flow-media sheets, 4 nylon peel plies, 2 layers of breather, and 16 layers of carbon fiber fabric. Through-thickness arrangement of these material layers was as follows: [1 aluminum mold + 1 layer of flow-media + 1 layer of nylon peel ply + 1 layer of breather + 8 layers of carbon fabric]$_S$. This arrangement was then wrapped with Stretchlon\textsuperscript{\textregistered} 800 bagging film and sealed with vacuum-sealant tape, ensuring space for both inlet and outlet connectors. 

Vinyl-ester resin was mixed with MEKP hardener in a container at a weight ratio of 100:1.25. The outlet of the vacuum bag was first connected to a vacuum pump to draw the bagging film to vacuum while clamping the inlet. The inlet of the vacuum bag was then submerged in the resin/hardener mixture to initiate the flow of the resin through the dry fabric layers, which was assisted by the vacuum created. It is important to note that the resin/hardener mixture was placed in a desiccator prior to infiltration in order to remove air bubbles from the mixture which may have formed during the mixing of resin and hardener. Upon completion of the resin transfer process, the laminate was cured at room temperature for 24 hours.

\subsubsection{Sample Dimensions}
Sixteen layers of dry carbon fabric resulted in panels with a nominal thickness of $\approx$ 4 $mm$. Composite test samples were water-jet cut to a length of 154 $mm$ and a width of 13 $mm$ from the panel. These dimensions were based on the ASTM standard D7264 \cite{ASTM1}, which requires a span-to-thickness ratio of 32:1 for accurate flexural results. 

\subsection{Sea Water Saturation}
Woven carbon/vinyl ester composite test samples were submerged in synthetic sea water  (Ricca Chemical - ASTM D1141 \cite{ASTM1141}) by placing inside a 20-gallon fish tank as per ASTM D5229 \cite{ASTM2}. 
In general for moisture uptake and diffusion studies, submerged test samples are periodically weighed to determine moisture uptake curves until moisture saturation is attained within the samples. Since the focus of this paper is not to determine moisture saturation content or diffusivity, the exposure time needed for the test samples to achieve saturation was chosen from a previous study conducted by the authors \cite{Garcia:2017}. Garcia et al. \cite{Garcia:2017} reported that these woven carbon/vinyl ester composite samples reach moisture saturation at approximately 120 days when exposed to synthetic sea water. The dimensions and materials used in the current study were identical to that used in Garcia et al. \cite{Garcia:2017}. Hence, test samples in the current study were submerged in synthetic seawater for a period of 140 days to achieve sea water saturation \cite{Alvarez:2007,SOBRINHO:2009,Koot:2001}.

It is important to note that the edges of the specimens were sealed with additional resin before exposing them to sea water to minimize infiltration through the edges. Sealing the edges with resin also considers certain real-life applications. For example, a ship hull made of composite-material panels would typically be in contact with sea water at the surfaces, and absorb the fluid only in the through-thickness direction. Composite panel surfaces in real applications are typically coated, however, they can be in contact with the environment in the case of coating wear-off or cracking. Thus, the approach adopted in this study enables the test samples towards one-dimensional absorption. Although, more detailed adaptations of real conditions such as the presence of coatings need to be studied for use in marine structures, they are beyond the scope of this paper. The samples were placed on supports within the fish tank to ensure exposure of the top and bottom faces to the surrounding medium.

\subsection{Mechanical Testing}
\begin{figure}[H]
\centering
\subfigure[]{
  \includegraphics[width=0.40\textwidth]{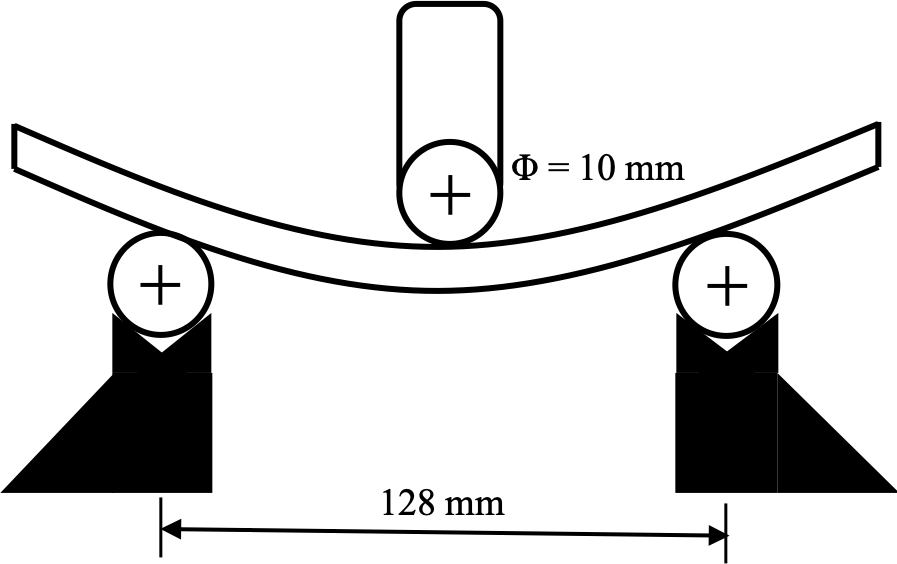} \label{threepointbend}
}
\centering
\subfigure[]{
  \includegraphics[width=0.5\textwidth]{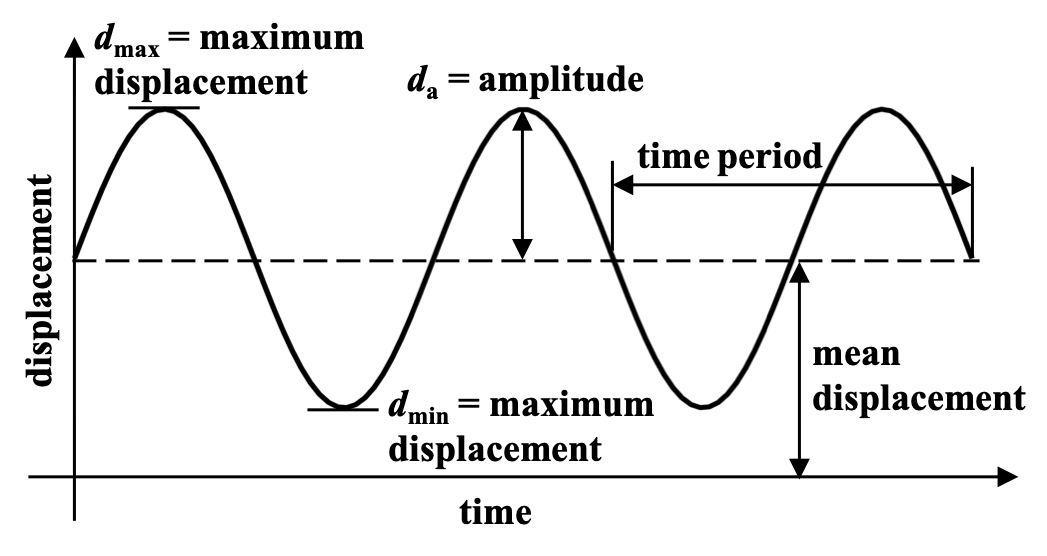}\label{sinewave}
}
\caption{(a) Three-point bend test setup; (b) Example of a displacement sine wave}
\end{figure}

Woven carbon/vinyl ester composite test samples were subjected to cyclic (fatigue) loading under flexure using the fixture shown in Fig. ~\ref{threepointbend}. The samples were tested under quasi-static three-point bend (flexure) loading first to determine the maximum mid-span deflection before failure, which averaged to 5.825 $mm$. The same fixture was used next to apply cyclic loading in flexure in a displacement-control setting to determine their flexural fatigue life. A sine waveform was used as the cyclic input displacement as shown in Fig. ~\ref{sinewave}. Four different displacement amplitudes ($2*d_a$) equal to 20\%, 40\%, 60\% and 80\% of maximum mid-span quasi-static deflection ($d_{max}^{st}$) and an amplitude ratio of $R = 0.1$ were chosen \cite{Slot:1969} for this study. The cyclic input displacement oscillates between maximum ($d_{max}$) and minimum ($d_{min}$) displacement such that $2*d_a = d_{max} - d_{min}$. The amplitude ratio ($R$) is defined as the maximum displacement over the minimum displacement, i.e. $R=\frac{d_{max}}{d_{min}}$.

Displacements (deflections) and loads were converted to stress and strain values using the relations in \ref{strainequation}, which are provided by the ASTM standard D7264 \cite{ASTM2}. 
\begin{equation}\label{strainequation}
\sigma=\frac{3PL}{2bh^2}; \quad \epsilon=\frac{6\delta h}{L^2}
\end{equation}

\noindent where, $\sigma$ is the stress at the outer surface at mid-span [MPa], $\epsilon$ is the strain at the outer surface [mm/mm], $P$ is the applied force [N], $\delta$ is the mid-span vertical deflection [mm], $L$ is the support span between the roller supports ($=128 mm$ here), $b$ is the specimen width [mm], $h$ is the specimen thickness [mm]. 
    
Two sets of test samples were considered in this study: dry and sea water saturated. 12 samples each for the two environmental conditions (dry and saturated) were tested under flexural fatigue loading. All tests were conducted at room temperature at a frequency of 1 Hz on an Instron 8801 servo-hydraulic machine. A Fast Track 8800 controller was used to create the above-mentioned sine wave function for each test, allowing the use of a close-loop technique that generated a feed-back signal which provided an instantaneous correction for small displacement amplitude deviations of the actuator. 

Fatigue process typically consists of three stages: (i) Initial fatigue damage leading to crack initiation, (ii) Progressive cyclic growth of crack, and (iii) Final fracture of the remaining cross-section \cite{Boyer}. For composites, due to complex internal architecture, different damage mechanisms manifest with progressive loading of the samples, which include matrix cracking and fiber breakage. In this study, these key damage mechanisms of fiber reinforced polymer composites were identified. The following procedure was followed for few samples to obtain images of damage mechanisms after varying cycles of loading. Towards that, the tests were periodically paused, upon which the samples were removed from the fixture and examined under a scanning electron microscope (SEM). The images were analyzed to identify the above-mentioned key damage mechanisms.
\subsection{Scanning Electron Microscopy (SEM)}

Different stages of fatigue damage were analyzed using a Zeiss LEO 1550VP (Carl Zeiss Jena GmbH - Planetariums, Germany) field emission scanning electron microscope (SEM). Sample’s cross-section was fine ground using SiC grinding paper up to 1200 grit size and then polished with 1-0.1 $\mu$m diamond solution on DACRON II Polishing Cloth (Pace Technologies, USA). The polished sample was mounted on to flexure fatigue fixture and periodically paused to analyze the cross-section under SEM and capture the damage modes at different stages. SEM samples were coated with gold using Hummer Jr Gold Sputter (Anatech,USA) under Argon atmosphere {(\em{70-80 mtorr}}) before loading into the SEM test chamber.  The gold coating was done to prevent the build-up of charges from the electrons absorbed by the specimen. After each fatigue test, the sample was mounted on a sample holder and placed into the exchange chamber. After that, the sample was transferred to the main working chamber. All the sample cross-sections were examined using a 15kV electron beam, and a secondary electron (SE) detector was used to acquire high-resolution images at different magnifications.

\section{Results and Discussion}\label{Results}

\subsection{Fatigue Test Results}
\begin{figure}[H]
  \centering
    \includegraphics[width=0.75\textwidth]{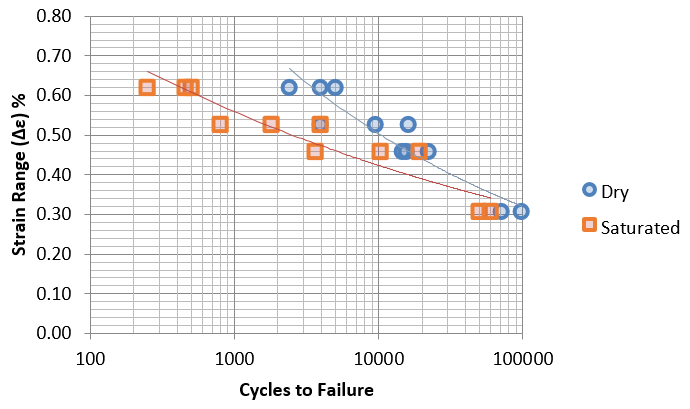}
    \caption{Strain-life fatigue curves for both dry and saturated specimens}\label{strainlifecurve}
\end{figure}

Fatigue failure of composites was assumed to occur when the load recorded at any time during the test dropped below 20\% of the average load of the first 10 cycles. The presentation of fatigue data for displacement-controlled tests commonly includes a graph of the percentage of total strain range ($\Delta\epsilon$ \%) versus the number of cycles to failure, also known as, the {\bf \em{strain-life curve}}. The strain-life curves for both dry and saturated samples are shown in Fig.~\ref{strainlifecurve}. Also, a summary of average cycles to failure for each test configuration is presented in Table \ref{fatiguetable}. 

From the strain-life curves shown in Fig.~\ref{strainlifecurve}, it is evident that sea water saturation in carbon/vinyl ester composites decreased the fatigue life of saturated specimens, approximately by up to 62\% across all strain ranges, when compared with dry samples. Reduction in fatigue life of these composites is attributed to the plasticization of vinyl ester matrix due to sea water saturation \cite{Mouritz2004,Sobrinho2009,Murthy2010,Visco2012,Yin2019}. The cycles to failure are comparable between dry and sea water saturated samples at lower strain ranges as is evident from Fig.~\ref{strainlifecurve}. However, the difference in the number of cycles to failure are drastically different at higher strain ranges, for example, at 0.62\%. This implies that the influence of sea water saturation on the fatigue life is more pronounced at higher strain levels. Hence, special considerations are warranted for designing structures made of woven carbon/vinyl ester composites expected to experience fatigue load with high strain range in marine environments.
\begin{table}[H]
\caption{{Matrix of fatigue testing and average results}}
  \centering
  \footnotesize
  \begin{tabular}{|p{2cm}|p{2cm}|p{4cm}|p{3cm}|}
    \hline 
Type & Strain Range ($\Delta\epsilon$) \% & Cycles to Failure $\pm$ Std. Dev. & \% avg. change with dry \\  \hline	
Dry  & 0.62 &  3830 $\pm$ 1110 & \\  	
Dry  & 0.53 &  5170 $\pm$ 3170 & \\  
Dry  & 0.46 &  17500 $\pm$ 3570 & \\ 
Dry  & 0.31 &   \textit{no failure}  & \\  
Saturated  & 0.62  & 400 $\pm$ 110 & -90 \\ 
Saturated  & 0.53  & 2170 $\pm$ 1290 & -59 \\
Saturated  & 0.46  & 10930 $\pm$ 6310 & -37.5 \\ 
Saturated  & 0.31  & \textit{no failure} & - \\ \hline
  \end{tabular}
  \label{fatiguetable}
\end{table}

\begin{figure}[H]
\centering
\subfigure[]{
\includegraphics[width=3.5in]{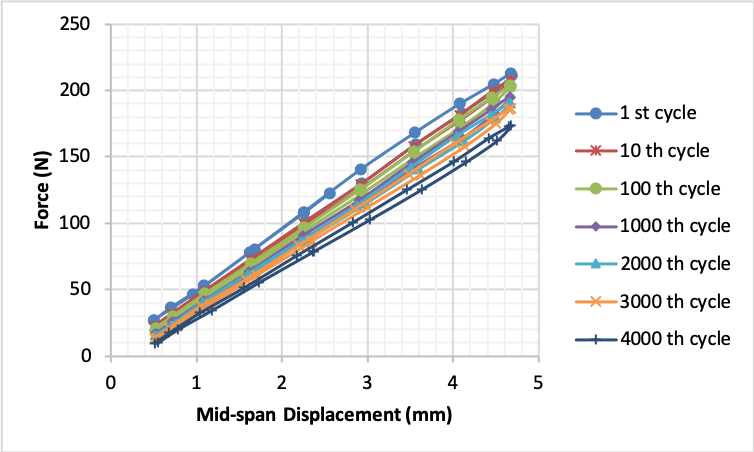}
}
\vspace{.1in}
\centering
\subfigure[]{
\includegraphics[width=3.5in]{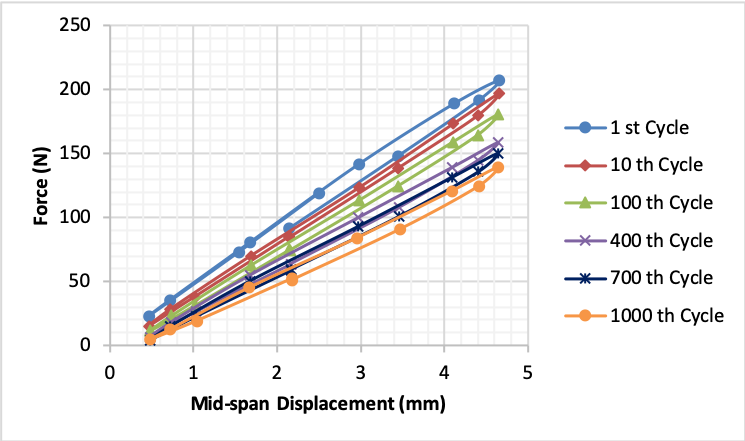}
}
\caption{(a) Load-displacement hysteresis loop for dry fatigued specimens under $\Delta\epsilon$ = 0.62\%; (b) Load-displacement hysteresis loop for saturated fatigued specimens under $\Delta\epsilon$ = 0.62\%}\label{hysteresis}
\end{figure}

Hysteresis loops were recorded in terms of load versus mid-span displacement values, which are typically used for establishing stiffness degradation and damage growth with increasing number of cycles. The hysteresis loops for both dry and saturated samples are shown in Fig.~\ref{hysteresis} (a-b) for percentage strain range ($\Delta\epsilon$) of 0.62\%. It is observed that the maximum load recorded for both types of conditioning reduced with increasing number of cycles. However, the dry samples (Fig.~\ref{hysteresis} (a)) showed a higher resistance to load reduction as compared to saturated samples (Fig.~\ref{hysteresis} (b)). For example, maximum load drops by approximately 33\% for saturated and 8\% for dry samples at the 1000$^{th}$ cycle at 0.62\% strain range. Therefore, higher damage is expected to be incurred in the saturated specimens. 

\begin{figure}[H]
\centering
\subfigure[]{
\includegraphics[width=3.5in]{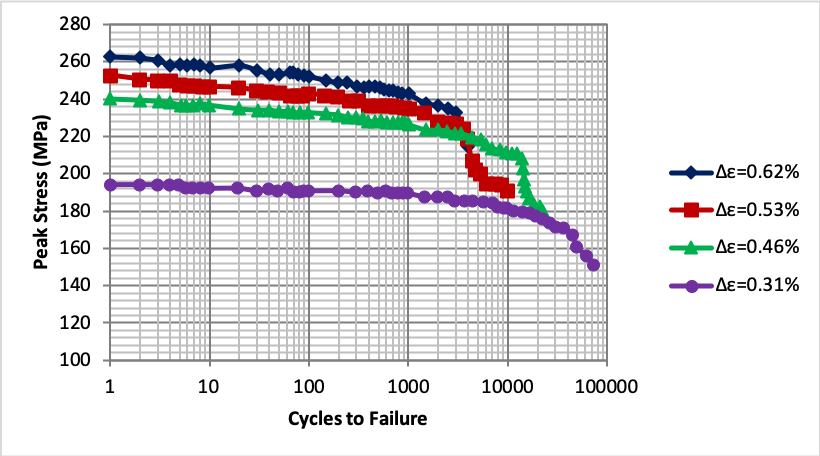}
}
\centering
\subfigure[]{
\includegraphics[width=3.5in]{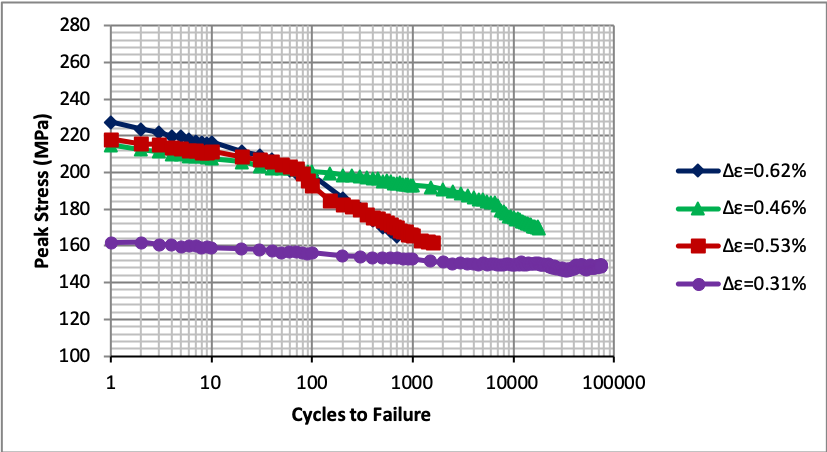}
}
\caption{Stress variation with increasing number of cycles of all strain range configurations for: (a) dry samples and (b) saturated samples}\label{stressvariation}
\end{figure}

The variation of peak stress versus number of cycles to failure \cite{Wu:2008} for each strain range is shown in Fig.~\ref{stressvariation} (a-b). These plots demonstrate that the cycles to failure drastically reduce due to sea water saturation for all strain ranges, i.e. average measured reduction between 37\% at 0.46\% strain range to 90\% at 0.62\% strain range. Hence, the influence of sea water saturation on the reduction of cycles to failure is higher (approximately 90\%) at higher strain ranges as shown in Fig.~\ref{stressvariation}(b). In addition, the influence of sea water saturation on the initial bending stiffness of these composites can be noted from Fig.~\ref{stressvariation}. A decreasing trend in initial bending stiffness is observed between dry and saturated samples from the stress values recorded at the 1$^{st}$ cycle for the same applied displacement, which are lower in Fig.~\ref{stressvariation} (b) than in Fig.~\ref{stressvariation} (a) for all strain percentages. 

\subsection{Damage modes}

\begin{figure}[H]
\centering
\subfigure[]{
\includegraphics[width=0.3\textwidth]{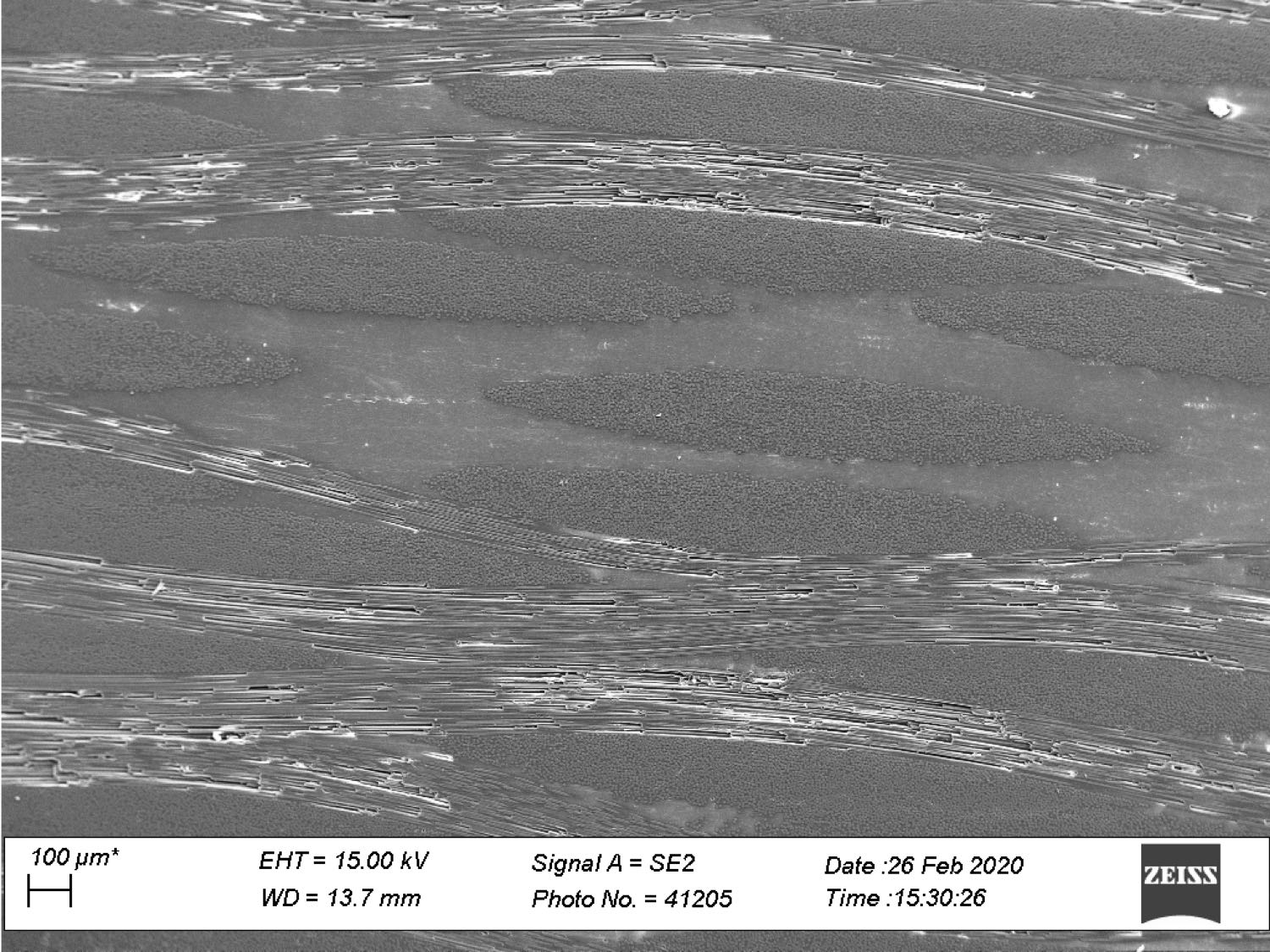}
\includegraphics[width=0.3\textwidth]{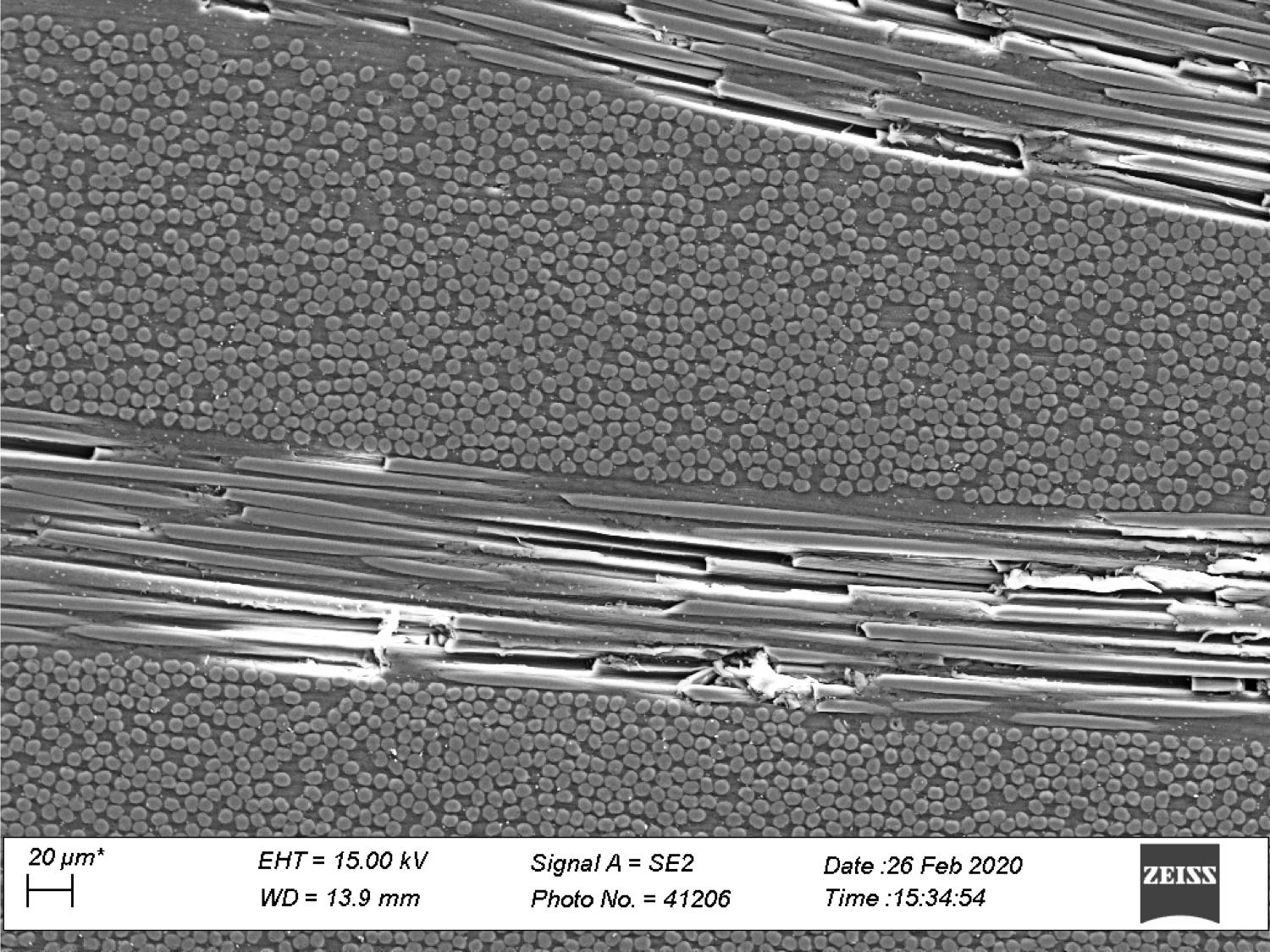}
\includegraphics[width=0.3\textwidth]{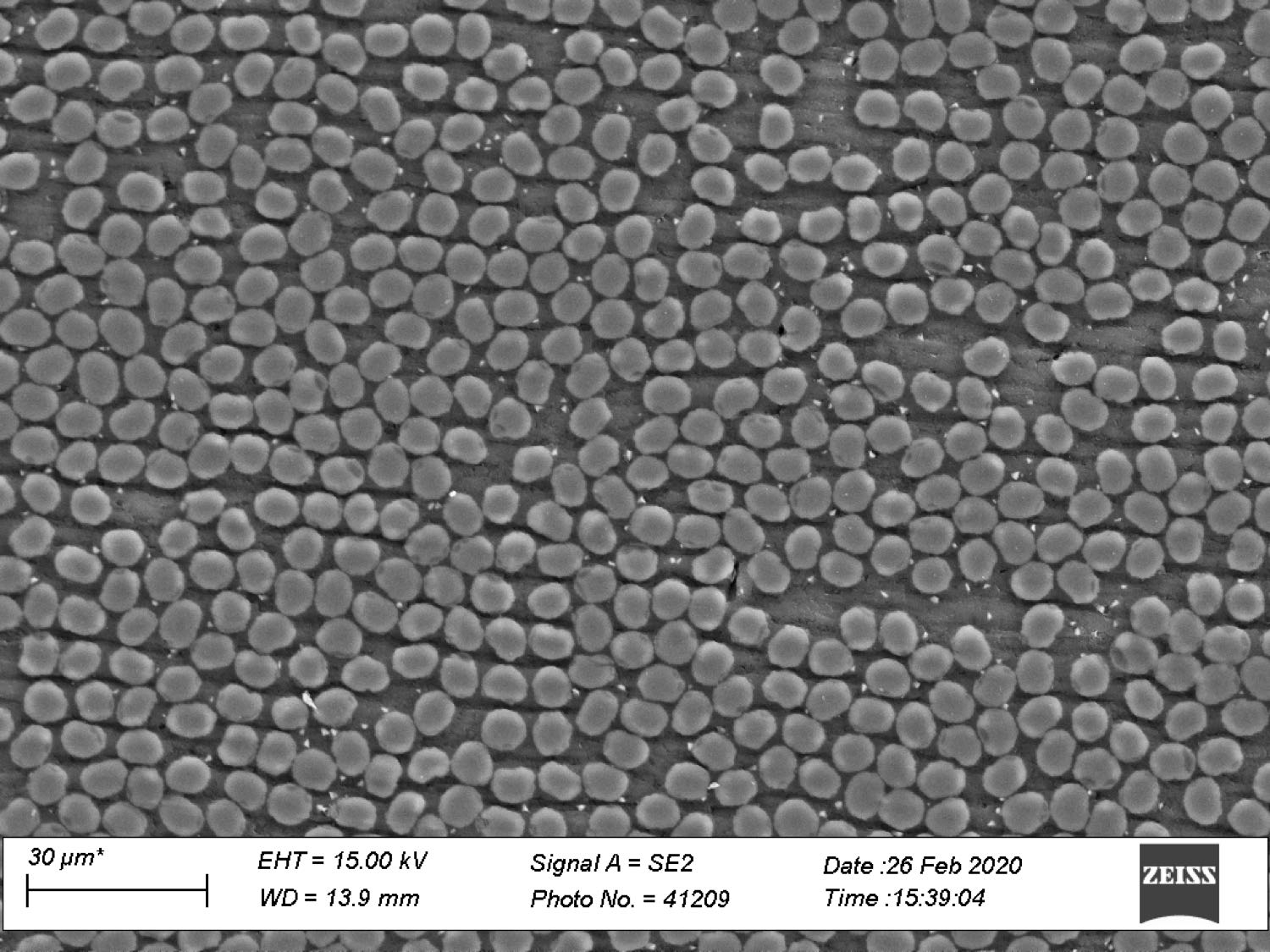}
}
\centering
\subfigure[]{
\includegraphics[width=0.3\textwidth]{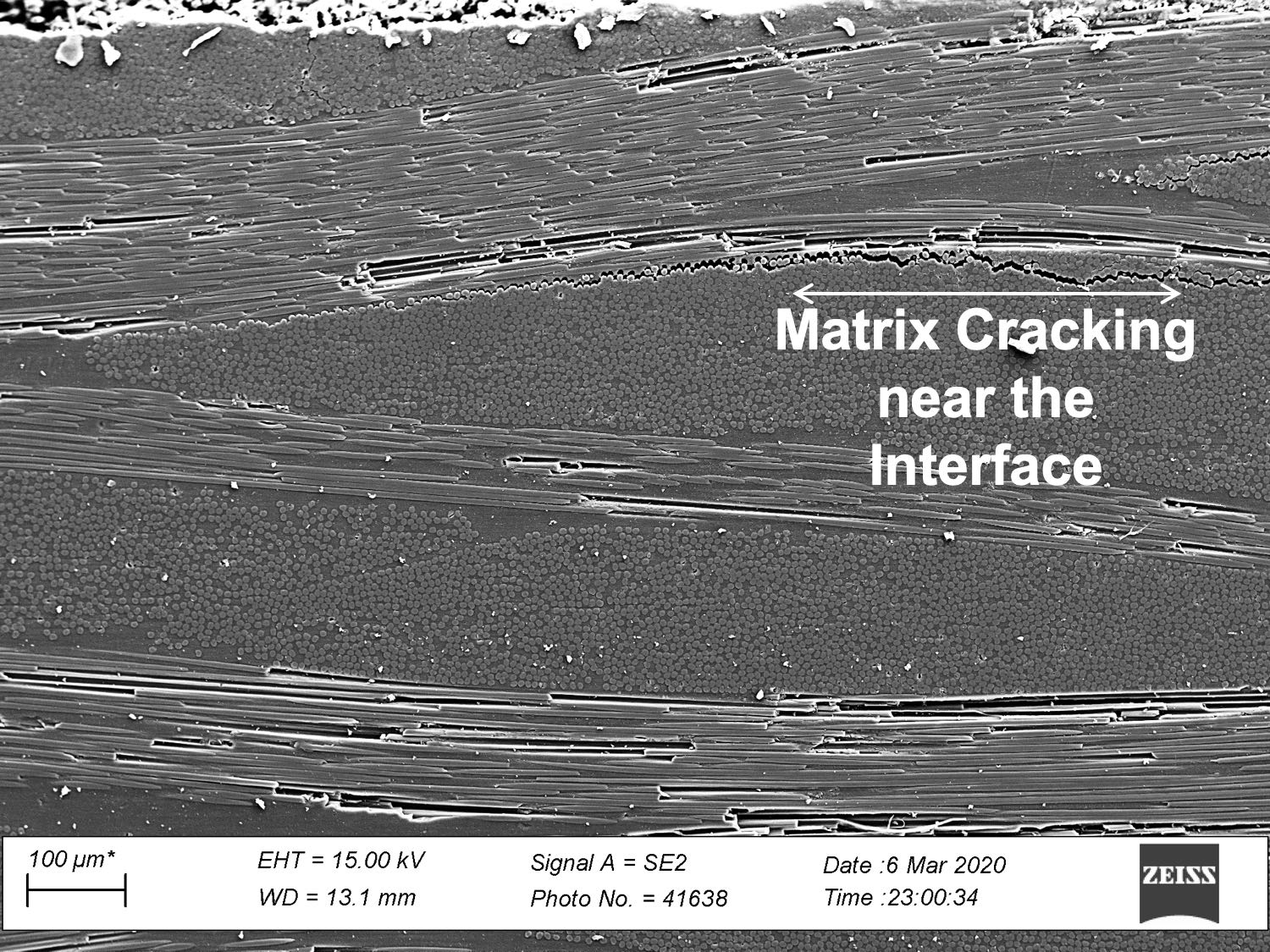}
\includegraphics[width=0.3\textwidth]{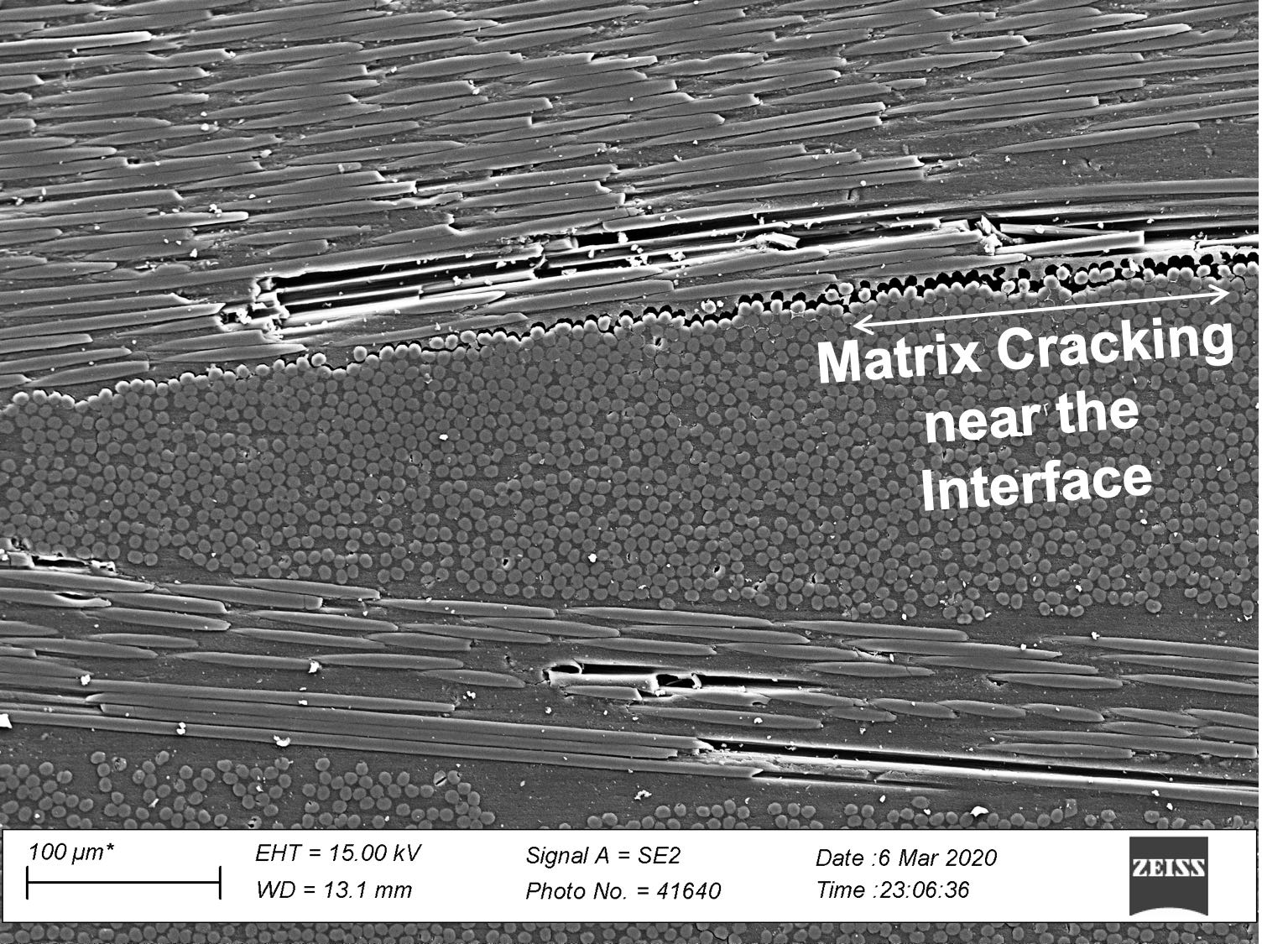}
\includegraphics[width=0.3\textwidth]{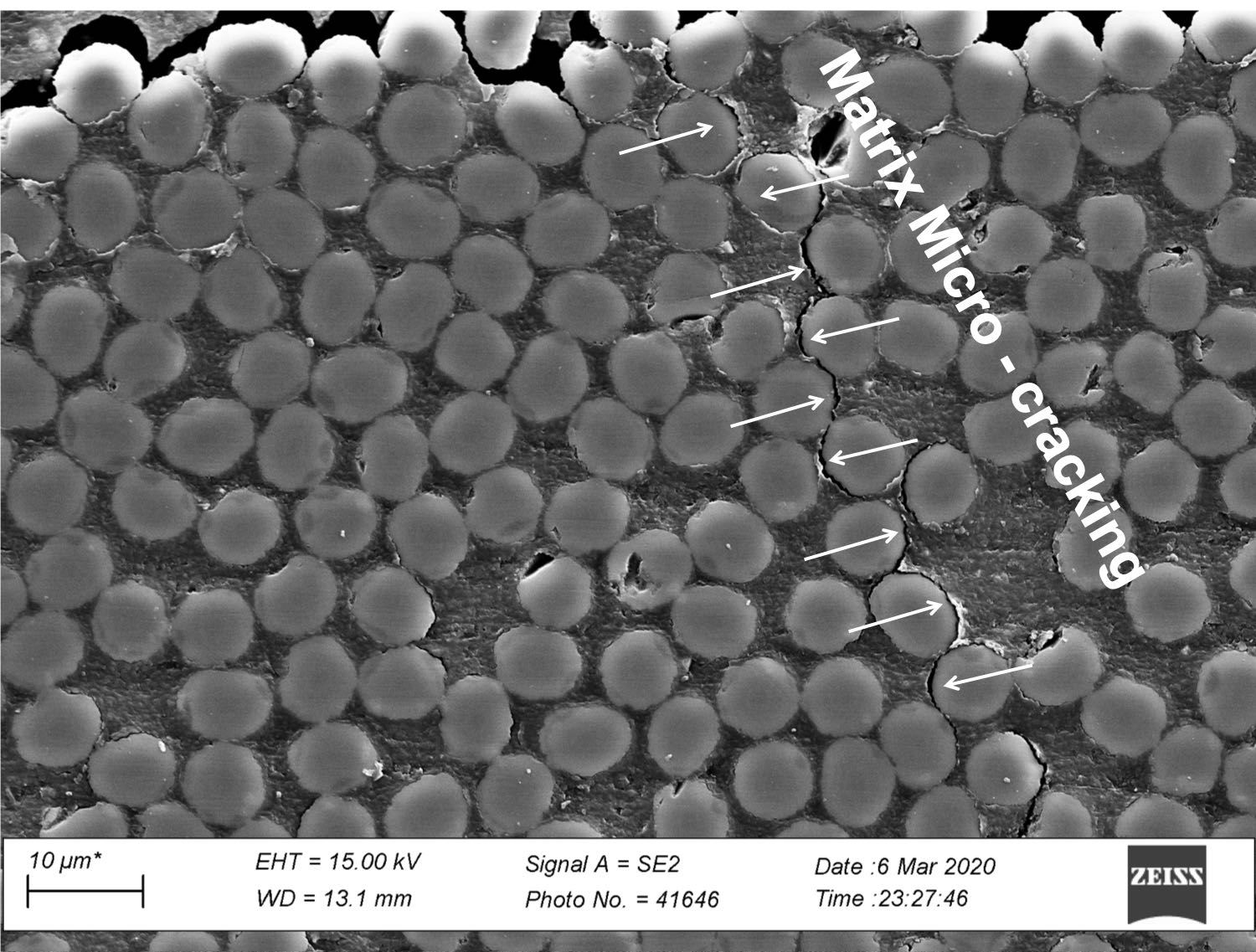}
}
\centering
\subfigure[]{
\includegraphics[width=0.3\textwidth]{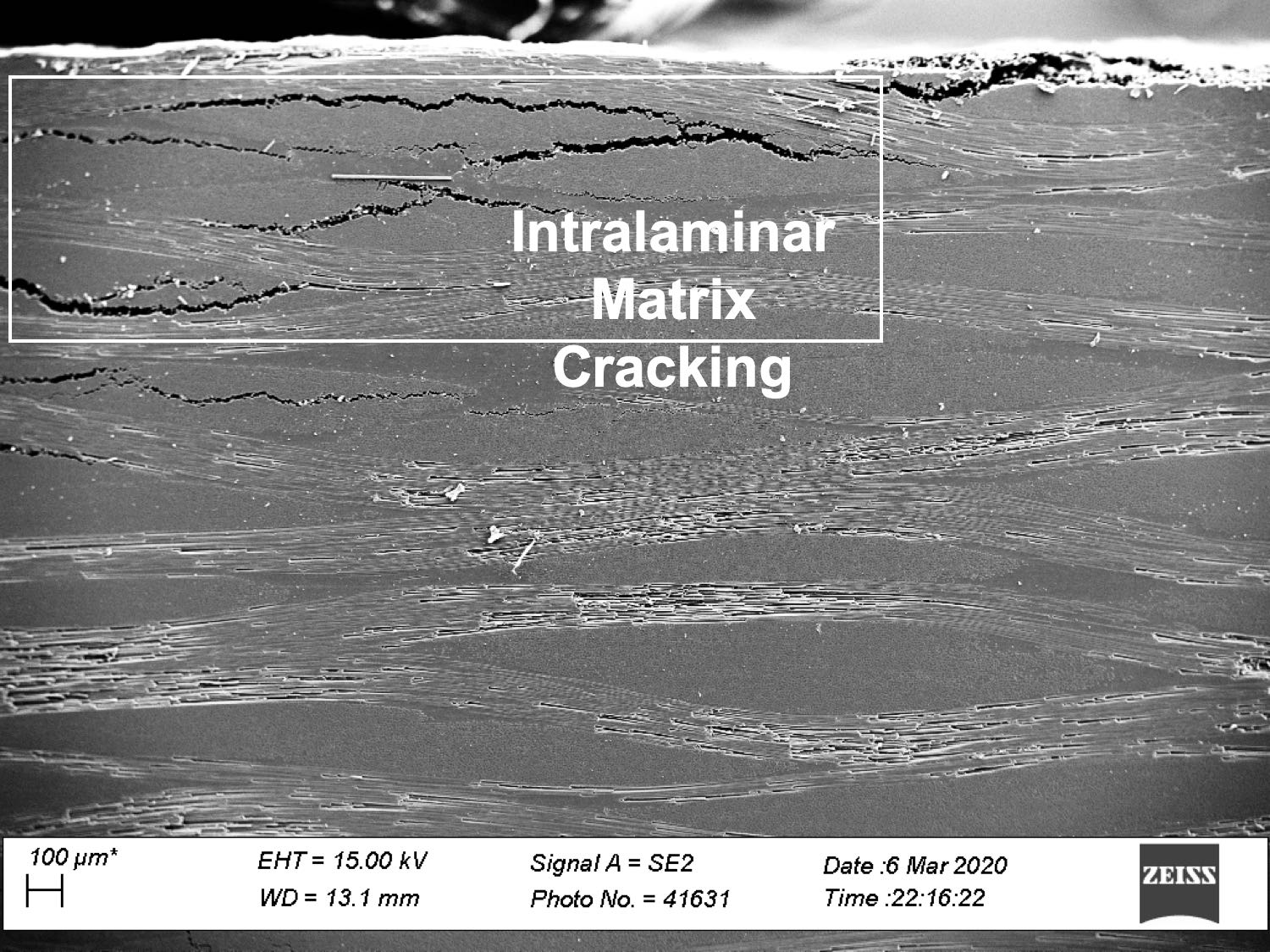}
\includegraphics[width=0.3\textwidth]{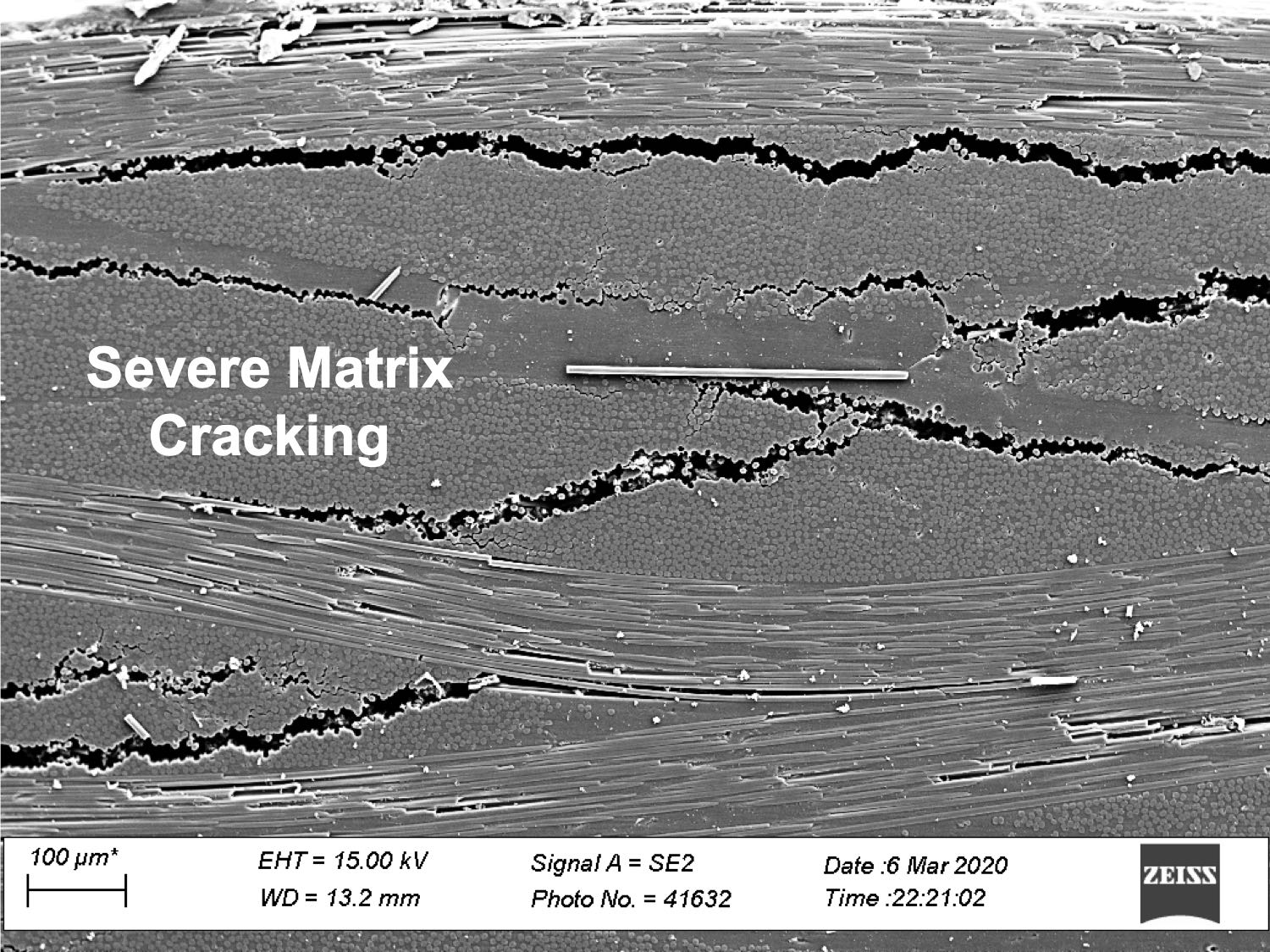}
\includegraphics[width=0.3\textwidth]{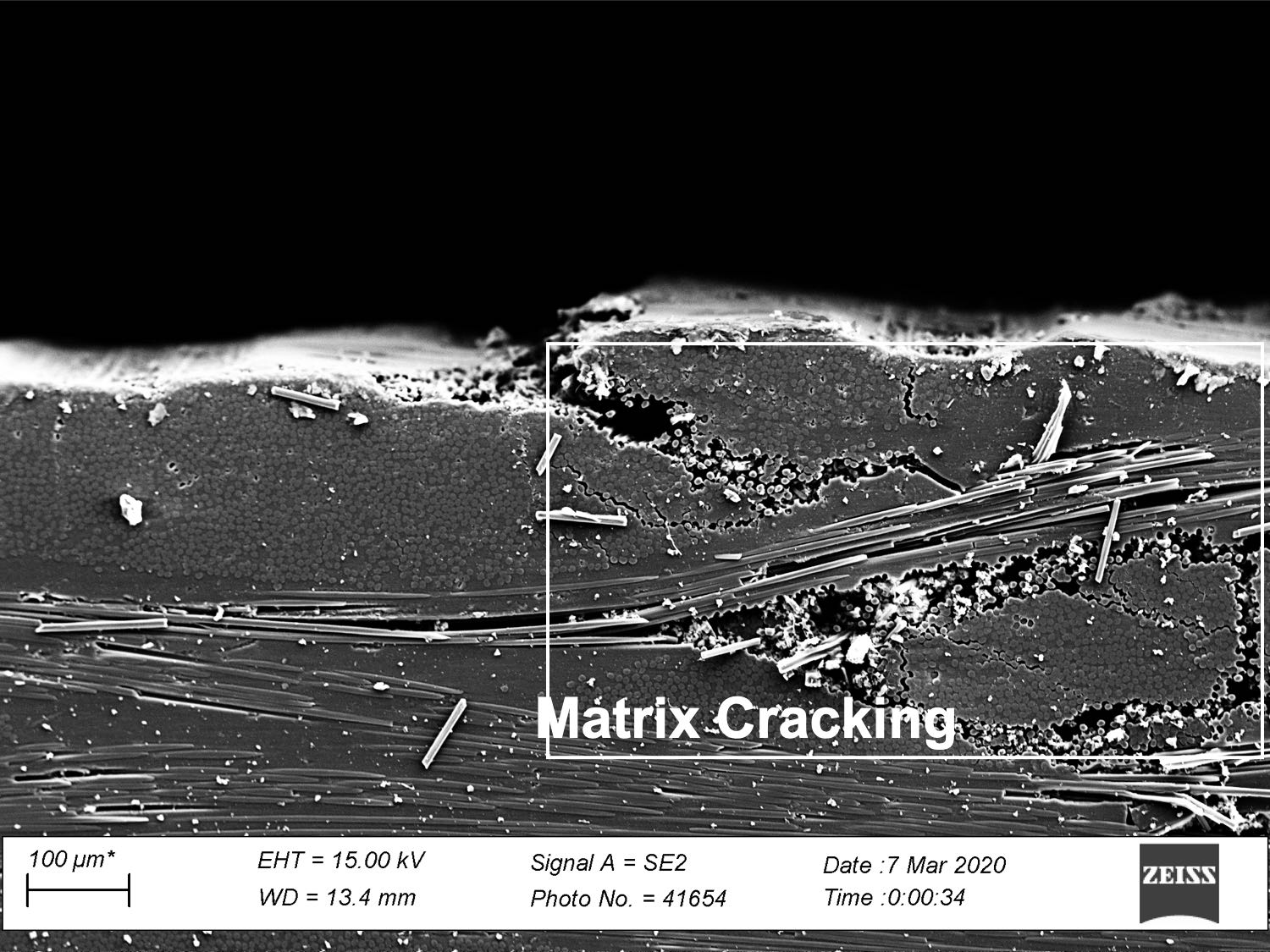}
} 
\centering
\subfigure[]{
\includegraphics[width=0.3\textwidth]{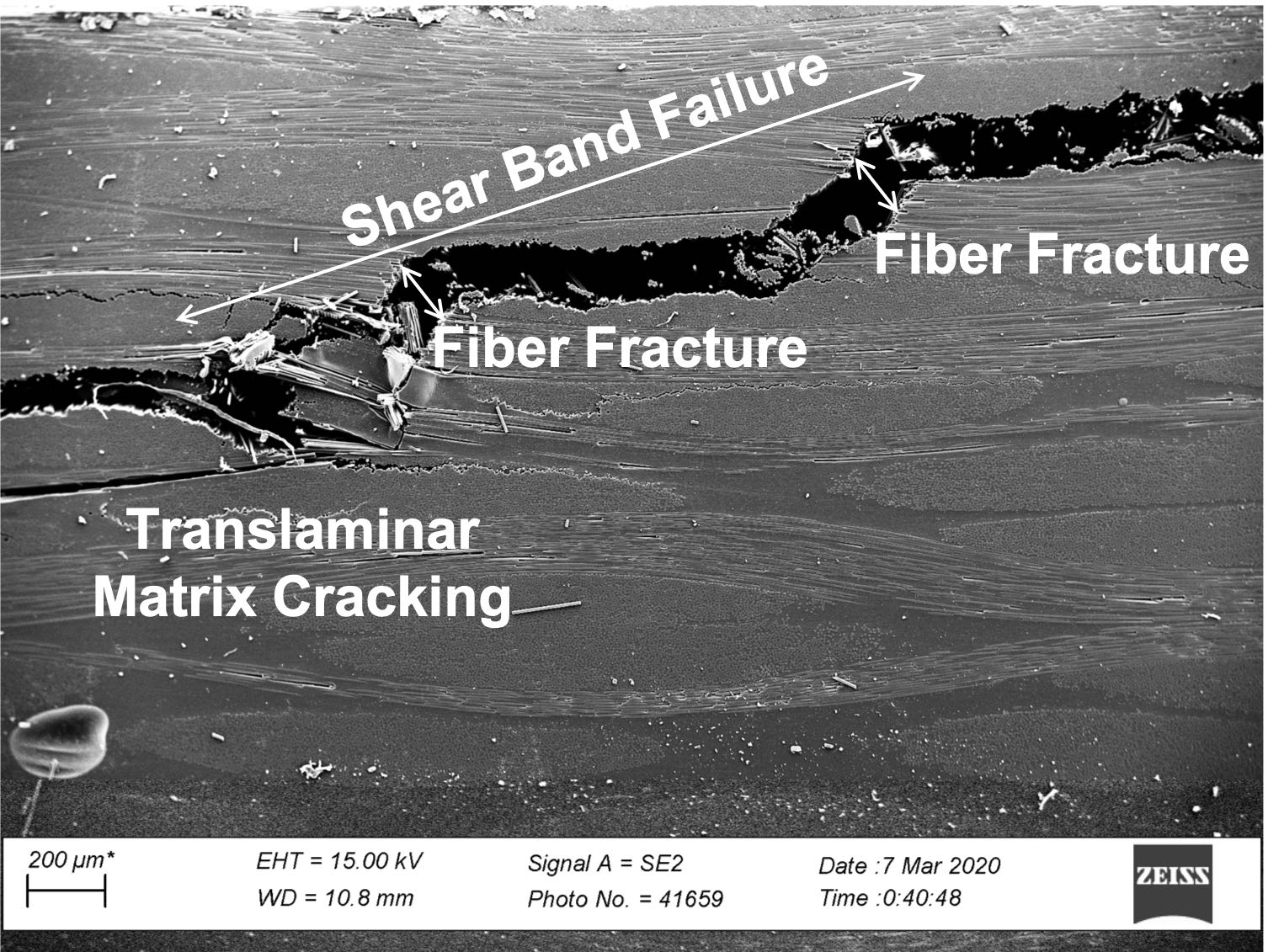}
\includegraphics[width=0.3\textwidth]{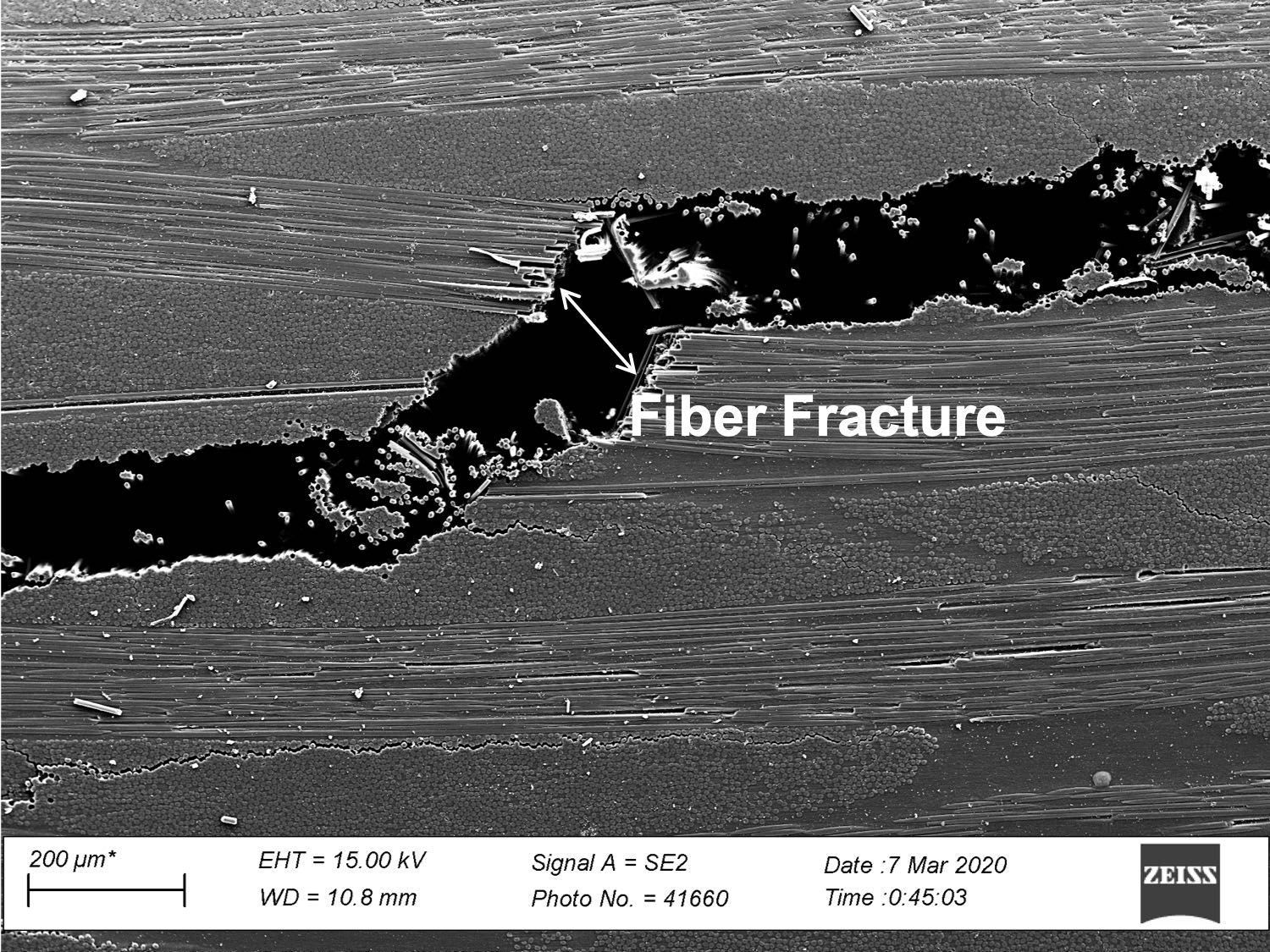}
\includegraphics[width=0.3\textwidth]{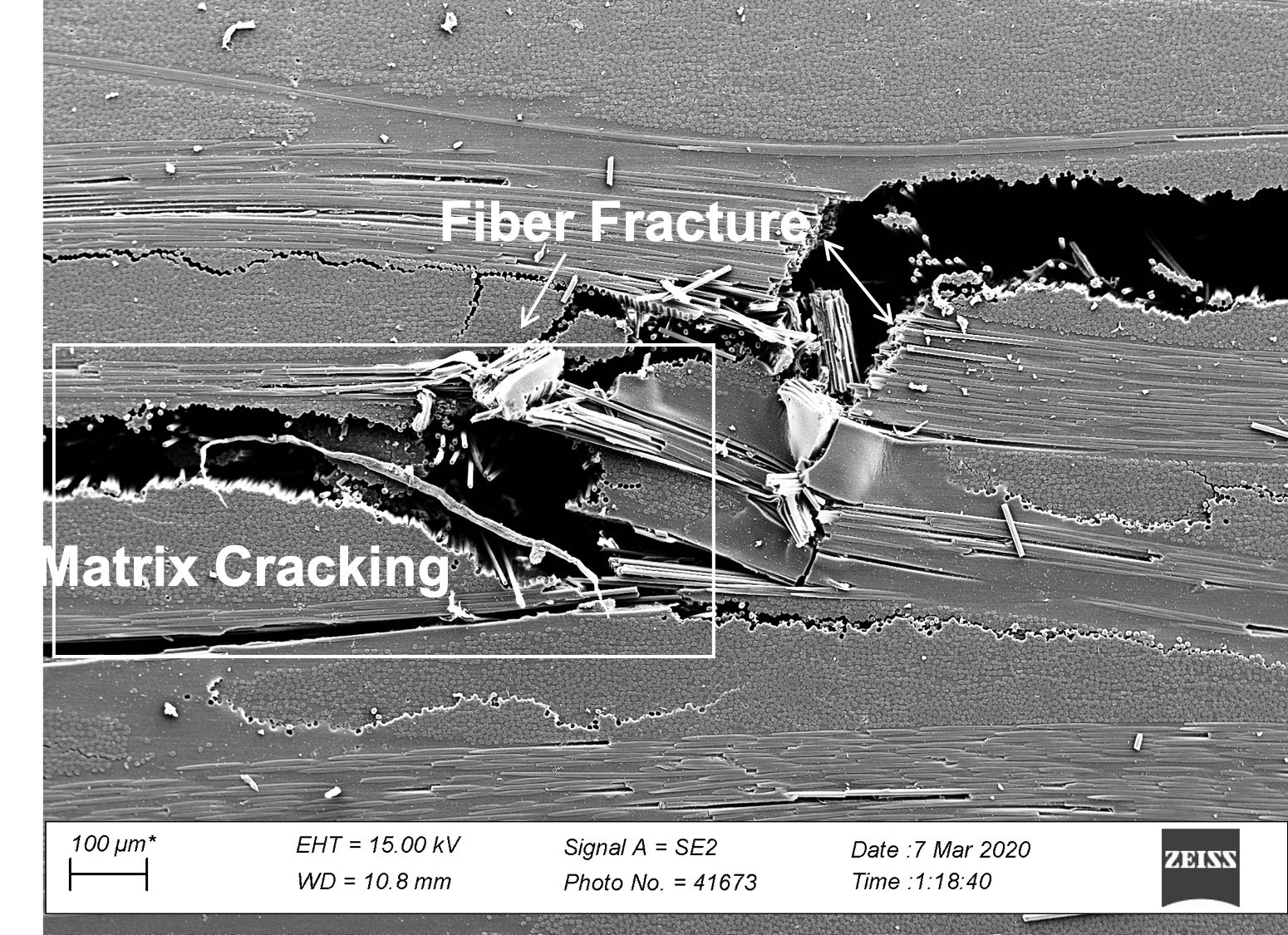}
}
\caption{Scanning electron microscopy images of composite microstructure: (a) pristine untested samples with no damage to serve as baseline for comparison (b) Stage 1: initiation of matrix cracking at the interface of finer bundles and matrix micro-cracking, (c) Stage 2: manifestation of several intralaminar matrix cracking, (d) Stage 3: shear band formation with fiber fracture and translaminar matrix cracking}\label{stagesdry}
\end{figure}

Typical damage modes that manifested under fatigue loading during the testing of dry specimens are shown in Fig.~\ref{stagesdry} (b-d). Fig.~\ref{stagesdry} (a) shows the microstructure of pristine untested sample that serve as baseline for comparison against other tested samples in Fig.~\ref{stagesdry} (b-d).
Stage 1 always initiated at the top surface of the specimen and manifested itself in the form of matrix cracking at the interface between bundles of fibers (tows). Stage 2 followed with a progressive growth of matrix cracking as intralaminar fracture. The rate of crack growth was dependent on the strain range applied, i.e., higher strain range percentages resulted in faster crack growth which manifested as reduced number of cycles to failure at higher strain ranges. Stage 3 ended the fatigue test by manifesting translaminar matrix crack extension along with fiber fracture, culminating with a fatal shear band failure mode in all cases. Saturated samples manifested similar stages of fatigue damage growth as that of dry samples. However, the saturated samples displayed lower resistance to intralaminar and translaminar fracture resulting in lower average number of cycles to failure, i.e. lower fatigue life.

\subsection{Fatigue Model}
Constant amplitude displacement-controlled flexural fatigue test data obtained for both dry and sea water saturated specimens shown in Figure~\ref{strainlifecurve} were used for determining their strain-life curves. The fatigue coefficient, $b$, and the fatigue exponent, $a$, are the intercept and slope of the best line fit to strain amplitude ($\Delta\epsilon$ \%) versus number of cycles to failure ($N_f$) data in log-log scale for the equation given by:

\begin{equation}\label{fitEqn}
    \Delta\epsilon = b (N_f)^a
\end{equation}

\begin{figure}[H]
\centering
\subfigure[]{
\includegraphics[width=0.47\textwidth]{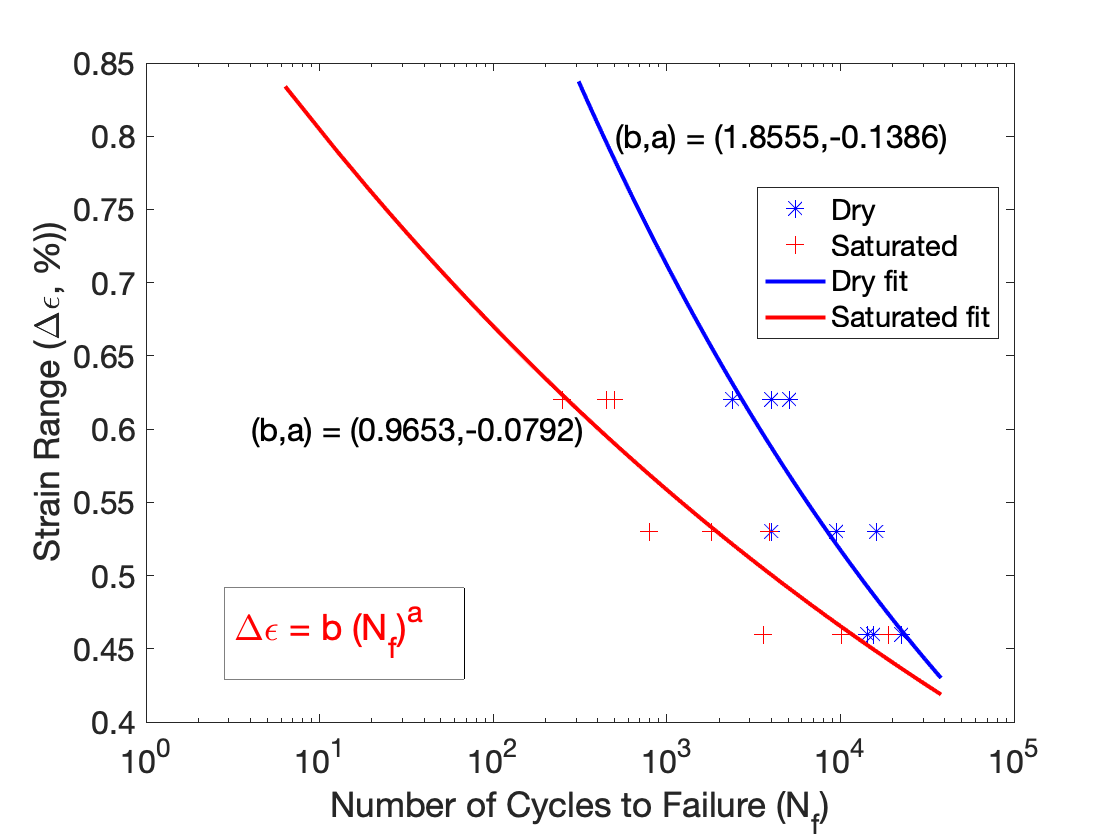}
}
\centering
\subfigure[]{
\includegraphics[width=0.47\textwidth]{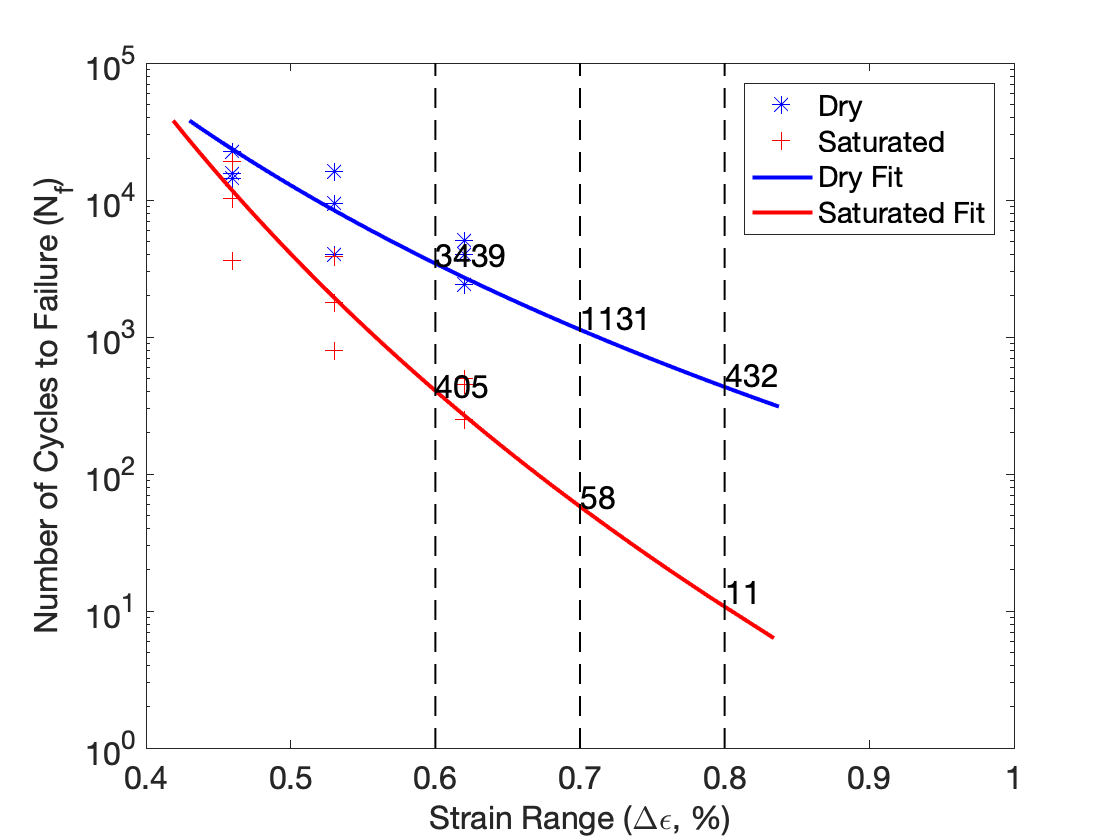}
} 
\caption{(a) Strain amplitude range versus number of cycles to failure, (b) Number of cycles to failure versus strain amplitude range }\label{fit}
\end{figure}

Figure~\ref{fit}(a) shows the fatigue life graphs for dry and sea water saturated composites with the non-linear fit given in Equation~\ref{fitEqn}. The fatigue coefficients and exponents for each case is also shown in Figure~\ref{fit}(a). The fatigue coefficients imply that the there is a significant reduction in the number of cycles to failure after sea water saturation indicated by a lower value of $b$. Further, the fatigue exponents imply that the difference in the number of cycles to failure reduce at lower strain amplitude ranges. It should be noted that the proposed model is valid for higher strain amplitude ranges above 0.4\%. This is because the specimens did not fail under fatigue at about $10^6$ cycles at a strain amplitude range of 0.31\%, and hence is not considered in this model.

Using this fitted model, the fatigue life in terms of number of cycles to failure ($N_f$) of both dry and sea water saturated composites can be determined as illustrated in Figure~\ref{fit}(b). The knockdown factors in cycles to failure for a range of strain amplitudes can be estimated using this model. As shown in Figure~\ref{fit}(b), the number of cycles to failure for a chosen strain range (say 0.6, 0.7 or 0.8\%) is the ordinates (y-axis intersection) of the dry (blue) and saturated (dry) fits. For example, at a strain range of 0.7\%, 1131 and 58 cycles to failure are estimated for dry and saturated cases, respectively. Also, as shown in the micro graphs in Figure~\ref{stagesdry}, the damage modes are driven by matrix cracking, which implies that the fatigue life depends on the fatigue behavior of vinyl ester resin. Hence, it is proposed that the fatigue model for the carbon woven/vinyl ester composite can be related to the fatigue response of vinyl ester in dry and sea water saturated conditions.

\section{Conclusions}

In this paper, the flexural fatigue life and damage mechanics of woven carbon/vinyl ester composites were characterized and compared under two different environmental conditions, which are dry and sea water saturated conditions (all at room temperature). Woven carbon/vinyl ester composite samples were tested using a standard three-point bend testing procedure by applying a displacement-controlled sinusoidal waveform at a frequency (f) of 1 Hz. A total of 12 specimens were tested for each type of environmental conditioning (i.e. dry and saturated). The saturated samples were exposed to synthetic seawater by submergence for a period of 140 days, which ensured moisture saturation. The data reported includes the strain-life curve, the load versus mid-span displacement hysteresis loops, and the peak stresses versus number of cycles to failure graphs. The following conclusions were made by the authors through the interpretation of the data presented:

\begin{itemize}
\item The fatigue life of sea water saturated samples decreased up to approximately 62\% as compared to dry samples. Synthetic seawater proved to be detrimental to the structural integrity of the composite with evidence of low resistance towards fatigue cracking and formation of shear band fracture caused by the plasticization of the vinyl ester resin. 

\item The cycles to failure are comparable between dry and sea water saturated samples at lower strain ranges, but are drastically different at higher strain ranges. That is, the average measured reduction is between 37\% at 0.46\% strain range to 90\% at 0.62\% strain range. This implies that the influence of sea water saturation on the fatigue life is more pronounced when the maximum cyclic displacement approaches maximum quasi-static deflection. Hence, special considerations are warranted for designing structures made of woven carbon/vinyl ester composites expected to experience fatigue load with high strain range in marine environments.

\item Hysteresis loops and peak stress variations with increasing number of cycles showed a monotonic decay in load for both dry and sea water saturated samples, where the latter displayed a higher rate of damage accumulation to failure than the former, thereby, resulting in lower flexural fatigue life. That is, maximum load drops by approximately 33\% for saturated and 8\% for dry samples at the 1000$^{th}$ cycle at 0.62\% strain range. 

\item For both dry and sea water saturated samples, key damage mechanisms observed include matrix crack initiation followed with a progressive growth manifested as intralaminar matrix cracking, and culmination with sample failure by fiber breakage and matrix shear band formation.

\item A non-linear fatigue model was fitted to the fatigue life curves of both dry and sea water saturated composites. The knock-down factors in cycles to failure for a range of strain amplitudes can be estimated using this model. These damage modes are driven predominantly by matrix cracking, which implies that the fatigue life depends on the fatigue behavior of vinylester resin. Hence, the proposed fatigue model can be used in the future for fatigue response of vinyl ester based composites in dry and sea water saturated conditions.

\end{itemize}

To summarize, this paper sheds light on the different predominant damages modes that influence the fatigue life of woven carbon fiber reinforced vinyl ester composites. Significant knockdown in fatigue life is observed at high strain ranges when saturated with sea water. Hence, special care should be taken when designing structures and components with woven carbon/vinyl ester composites in applications where cyclic loading and sea water exposure are commonplace, as these conditions can induce premature damage in the microstructure of the composite and ultimately cause catastrophic failure of a structure.

\section*{Author Contributions}
The first two authors contributed equally to this paper. Author P.P. contributed to the conceptualization, methodology, writing -- review and editing, visualization, verification, supervision, project administration and funding acquisition. Author R.G. contributed to the methodology, formal analysis, investigation and writing -- original draft preparation. Authors M.I. and V.D. contributed towards revising the paper and add valuable information to the paper.

\section*{Funding}
This research was partly funded by U.S. Department of Defense (DoD) Office of Naval Research Young Investigator Program (ONR-YIP) Grant [N00014-19-1-2206] through {\em{Sea-based Aviation: Structures and Materials Program}} and DoD HBCU/MI Basic Research Grant [W911NF-15-1-0430].

\section*{Acknowledgment}
The authors would like to acknowledge the technical support by Eduadro Garcia, a graduate student at the University of Texas at El Paso.
 


\bibliographystyle{unsrt}
\bibliography{ref_pp}

\end{document}